\documentclass[prl,reprint,showpacs,groupedaddress]{revtex4-1}
\usepackage{amsmath,amssymb,amsfonts}
\usepackage{graphicx}
\usepackage{txfonts}
\usepackage{color}
\usepackage{xspace}

\newcommand{\ket}[1]{{|#1\rangle}}
\newcommand{\eqnrefp}[1]{{[Eq.~(\ref{#1})]}}
\newcommand{\eqnreft}[1]{{Eq.~(\ref{#1})}}

\newcommand{\sech}[1]{\mathrm{sech}\left( #1 \right)}
\newcommand{\figreft}[2]{Fig.~\ref{#1}#2}
\newcommand{\figreftcap}[2]{Fig.~\ref{#1}#2}
\newcommand{\figrefp}[2]{[Fig.~\ref{#1}#2]}
\newcommand{\BSW}{BSW\xspace}
\newcommand{\BSWs}{BSWs\xspace}

\newcommand{\red}[1]{\textcolor{red}{#1}}
\newcommand{\green}[1]{\textcolor{green}{#1}}
\newcommand{\blue}[1]{\textcolor{blue}{#1}}
\newcommand{\magnta}[1]{\textcolor{magenta}{#1}}
\newcommand{\cyan}[1]{\textcolor{cyan}{#1}}

\begin{document}

\title{Realizing bright matter-wave soliton collisions with controlled relative phase}

\author{T. P. Billam}
\email{t.p.billam@durham.ac.uk}
\author{S. L. Cornish}
\author{S. A. Gardiner}
\affiliation{Department of Physics, Durham University, Durham DH1 3LE, United Kingdom}

\date{\today}

\begin{abstract}
We propose a method to split the ground state of an attractively interacting
atomic Bose-Einstein condensate into two bright solitary waves with controlled
relative phase and velocity. We analyze the stability of these waves against
their subsequent re-collisions at the center of a cylindrically symmetric,
prolate harmonic trap as a function of relative phase, velocity, and trap
anisotropy. We show that the collisional stability is strongly dependent on
relative phase at low velocity, and we identify previously unobserved
oscillations in the collisional stability as a function of the trap anisotropy.
An experimental implementation of our method would determine the validity of
the mean field description of bright solitary waves, and could prove an
important step towards atom interferometry experiments involving bright
solitary waves.
\end{abstract}

\pacs{
03.75.Lm,    
05.45.Yv,    
37.25.+k     
}

\maketitle

Bright solitary waves (\BSWs) in attractively-interacting atomic Bose-Einstein
condensates (BECs) are an intriguing example of a nonlinear wave phenomenon in
a degenerate quantum gas \cite{khaykovich_etal_science_2002,
strecker_etal_nature_2002, cornish_etal_PRL_2006}. In the mean-field,
Gross-Pitaevskii equation (GPE) description, \BSWs in a quasi-1D BEC with no external
(axial) trapping potential correspond exactly to bright solitons in the focusing nonlinear
Schr\"{o}dinger equation (NLSE)
\begin{equation}
i \frac{\partial \psi(x,t)}{\partial t} = \left[ -\frac{1}{2}
\frac{\partial^2}{\partial {x}^2} - \vert \psi(x,t)\vert^2 \right] \psi(x,t).
\label{nlse}
\end{equation}
This integrable equation describes a diverse range of physical systems in
addition to BECs \cite{dauxois_book}, and its bright soliton solutions have
been extensively studied in the context of nonlinear optics \cite{*[] [{ [Sov.
Phys. JETP {\bf 34}, 62 (1972)]}] zakharov_shabat_1972_russian,
gordon_opt_lett_1983, satsuma_yajima_1974, kodama_hasegawa_1991,
afanasjev_vysloukh_j_opt_soc_am_b_1994}. The addition of a harmonic trapping
potential to \eqnreft{nlse} leads to non-integrability, but \BSWs in
harmonically trapped, quasi-1D BECs remain, in the GPE description, highly
soliton-like; they collide elastically over a large parameter regime, and their
asymptotic trajectories follow particle models applicable to solitons
\cite{martin_etal_prl_2007, *martin_etal_PRA_2008, poletti_etal_prl_2008}.
Relaxing the quasi-1D restriction reduces the soliton character of \BSWs
further, but \BSWs in the 3D GPE description retain many soliton-like
characteristics \cite{parker_etal_JPB_2008, *parker_etal_physicaD_2008},
including an absence of dispersion and the existence of a {\it well-defined relative
phase between \BSWs}.  If actual 3D \BSWs possess a well-defined relative phase
when realized experimentally, one can envisage a \BSW interferometer, akin to
current matter-wave interferometers but leveraging the small size, coherence,
and non-dispersive nature of \BSWs \cite{strecker_etal_nature_2002}.  Indeed,
\BSWs have already been proposed as a metrological tool for the study of
atom-surface interactions \cite{cornish_etal_physicad_2009}.

Experiments to date have produced both individual
\cite{khaykovich_etal_science_2002} and multiple
\cite{strecker_etal_nature_2002, cornish_etal_PRL_2006} \BSWs as remnants from
the collapse \cite{donley_etal_nature_2001, *pitaevskii_pla_1996,
parker_etal_JPB_2007} of a larger condensate.  These \BSWs were capable, in the
case of multiple \BSWs, of surviving many mutual re-collisions at the trap
center \cite{strecker_etal_nature_2002, cornish_etal_PRL_2006}. In the quasi-1D
regime the observed \BSW motion matches the GPE description of \BSWs with
relative phase $\Phi=\pi$ \cite{strecker_etal_nature_2002,
al_khawaja_etal_PRL_2002, *strecker_etal_NJP_2003}. In the 3D regime, however,
\BSWs are not universally stable against multiple re-collisions; numerical
integration of the GPE reveals that slow 3D \BSWs retain their form for fewer
collisions when their relative phase, $\Phi$, is equal to $0$ than when
$\Phi=\pi$ \cite{parker_etal_JPB_2008, parker_etal_physicaD_2008}.  The long
lifetimes of 3D \BSWs seen in experiment thus seem to imply that their relative
phase $\Phi=\pi$ \cite{cornish_etal_PRL_2006, parker_etal_JPB_2008,
parker_etal_physicaD_2008}.  Modulational instability and the shorter lifetime of
colliding 3D \BSWs when $\Phi=0$ have been identified as contributory causes
to these apparent anti-phase relations in both regimes
\cite{al_khawaja_etal_PRL_2002, strecker_etal_NJP_2003, carr_brand_PRL_2004,
parker_etal_JPB_2008, parker_etal_physicaD_2008}. However, recent simulations
of \BSW collisions incorporating quantum noise have been interpreted as showing
the dynamics and collisional stability of \BSWs to be {\it phase-independent},
with the dynamics for {\it all} relative phases corresponding to the GPE
description for $\Phi=\pi$ \cite{davis_group_njp_2009}.  Furthermore, no 3D GPE
simulation of the collapse process has produced \BSW remnants matching those
observed in experiment \cite{davis_group_njp_2009}.  These considerations leave
the question open: {\it are experimentally observed atomic \BSWs well-described
by an effective single-particle wavefunction, propagated by the GPE?}

In this letter we propose an experiment to answer this question. We describe a
method which, in a velocity- and phase-controlled way, splits a single \BSW in
an axisymmetric harmonic trap into two outgoing \BSWs which repeatedly
re-collide at the trap center. Using the GPE, we demonstrate that such pairs of
\BSWs can be realized in an atomic BEC, and analyze their dynamics and
collisional stability. We explore the crossover from quasi-1D to fully 3D
regimes, examining the effects of velocity and relative phase on the number of
collisions for which the \BSWs remain soliton-like, $C_{\rm 1D}$. In addition
to demonstrating the expected $C_{1 \rm D}$ phase dependence for slow
collisions, we show that $C_{\rm 1D}$ has a strong oscillatory dependence on
the trap anisotropy, due to resonances between the frequency of the \BSWs'
radial oscillations and the frequency with which the two \BSWs collide. The
experimental presence (absence) of the predicted phase and anisotropy
dependencies would indicate the (in)sufficiency of the GPE description of
\BSWs; either of these outcomes would be an important result. In particular,
sufficiency of the GPE description implies a well-defined relative phase
between \BSWs, paving the way for future atom interferometry experiments using
\BSWs, towards which our proposed \BSW generation method would represent an
important step.

We begin with the GPE for a BEC of $N$ atoms of mass $m$ and (attractive)
$s$-wave scattering length $a_s<0$, held within a cylindrically symmetric,
prolate harmonic trap, \begin{equation} i\hbar\frac{\partial
\Psi(\mathbf{r},t)}{\partial t} = \left[-\frac{\hbar^2}{2m} \nabla^2 +
V(\mathbf{r}) - |g_{\rm 3D}|\vert \Psi(\mathbf{r},t) \vert^2 \right]
\Psi(\mathbf{r},t), \label{3dgpe_full_units} \end{equation} where $g_{\rm 3D} =
4\pi N a_s \hbar^2 / m$, the condensate wavefunction $\Psi(\mathbf{r},t)$ is
normalized to one, and $V(\mathbf{r}) = m[\omega_x^2 x^2 + \omega_r^2
(y^2+z^2)]/{2}$, where $\omega_x$ and $\omega_r>\omega_x$ are the axial and
radial trap frequencies. With strong radial confinement the system is quasi-1D
and can be described by \eqnreft{3dgpe_full_units} with $\Psi(\mathbf{r},t)
\rightarrow \Psi(x,t)$, $V(\mathbf{r}) \rightarrow m\omega_x^2 x^2 / 2$, and
$g_{3 \rm D} \rightarrow g_{1 \rm D} = 2 \hbar \omega_{r} |a_{s}| N$
\cite{Note_1D_reduction}. The resulting configuration has two key length scales; the
harmonic length $a_0 = \sqrt{\hbar/m\omega_x}$, and the soliton length $b_0 =
\hbar^{2} / m g_{1\mathrm{D}}$. Rescaling all lengths to units of $b_0$ and all
times to units of $\hbar^3 / m g_{1 \rm D}^2$ \cite{martin_etal_PRA_2008}
produces the dimensionless 1D GPE
\begin{equation} i \frac{\partial \psi(x,t)}{\partial t} = \left[
-\frac{1}{2} \frac{\partial^2}{\partial {x}^2} + \frac{\omega^2 x^2}{2} - \vert
\psi(x,t)\vert^2 \right] \psi(x,t), \label{1dgpe_soliton_units} 
\end{equation}
where $\omega = (b_0 / a_0)^2$ is a dimensionless effective trap strength
\cite{Note_1D_normalization}. In this letter, we consider the effects of
abruptly increasing the scattering length magnitude in such a BEC from initial
$a_{s}^{0}$ to $a_{s} = \alpha^{2}a_{s}^{0}$ ($\alpha > 1$ and $a_{s},
a_{s}^{0} < 0$), with initial condition
\begin{equation}
\label{modulated_ic}
\psi(x,t=0) = \psi_0(x) = \psi_\alpha(x) \cos \left( \frac{k x}{2\alpha^2} +\frac{\Phi}{2}
\right), 
\end{equation}
where $\psi_\alpha(x)$ is the \BSW ground state of the BEC for scattering
length $a_s^0$. We first consider the quasi-1D limit, where a stable ground
state $\psi_\alpha(x)$ always exists, and then 3D, where the stable ground
state $\psi_\alpha(\mathbf{r})$ exists only for $|a_s^0|$ below the critical
value for the onset of collapse, $|a_s^{\rm c}|$ \cite{parker_etal_JPB_2007}. 

The ground state $\psi_\alpha(x)$ may be made by using a magnetic Feshbach
resonance to adiabatically change the scattering length from being initially
repulsive to a negative value, $a_s^0$, with $|a_s^0|<|a_s^{\rm c}|$. The rapid
change from $a_s^0$ to $a_s=\alpha^2 a_s^0$ could then exploit the same
resonance.  The density modulation that transforms $\psi_{\alpha}(x)$ into
$\psi_{0}(x)$ may be achieved with a second internal atomic state in an
interference protocol: writing the total state of the condensed atoms as
$\psi_{+}(x)|+\rangle + \psi_{-}(x)|-\rangle$, we begin with all atoms in
internal state $|+\rangle$ [i.e., $\psi_{+}(x) = \psi_{\alpha}(x)$ and
$\psi_{-}(x)=0$].  Applying a resonant $\pi/2$ pulse to the internal state
transition yields $\psi_{+}(x) = \psi_{-}(x) = \psi_{\alpha}(x)/\sqrt{2}$.  We
now imprint equal and opposite momenta on the two internal states, giving
$\psi_{\pm}(x) =  \exp[\pm i (Kx +\Phi)/2] \psi_\alpha(x) / \sqrt{2}$, which is
then transformed into
\begin{equation}
\begin{split}
\psi_+(x) &= \cos \left[ (Kx +\Phi)/2 \right] \psi_\alpha(x), \\
\psi_-(x) &= i\sin \left[ (Kx +\Phi)/2 \right] \psi_\alpha(x),
\end{split}
\end{equation}
by a second $\pi/2$ pulse. Using resonant light to rapidly expel atoms in state
$\ket{-}$ from the trap leaves those in state $\ket{+}$ with wavefunction
$\psi_0(x)$ \eqnrefp{modulated_ic}, with $k = \alpha^2 K$, and with $\Phi$
determined by the phase accumulated at the center of the \BSW. Note that the
loss of atoms between $\psi_\alpha(x)$ and $\psi_0(x)$ %
is balanced by the change in normalization; $N$ denotes the
\textit{initial} atom number. There are many potential implementations of this
protocol. Using a two-component GPE, we have simulated an implementation that
uses $^{85}$Rb atoms in the quasi-1D regime with an applied magnetic field
gradient to transfer momentum: we find this simple prototype to be capable of
generating initial conditions close to \eqnreft{modulated_ic} using current
experimental technology \cite{Note1}.

\begin{figure}[t]
\includegraphics{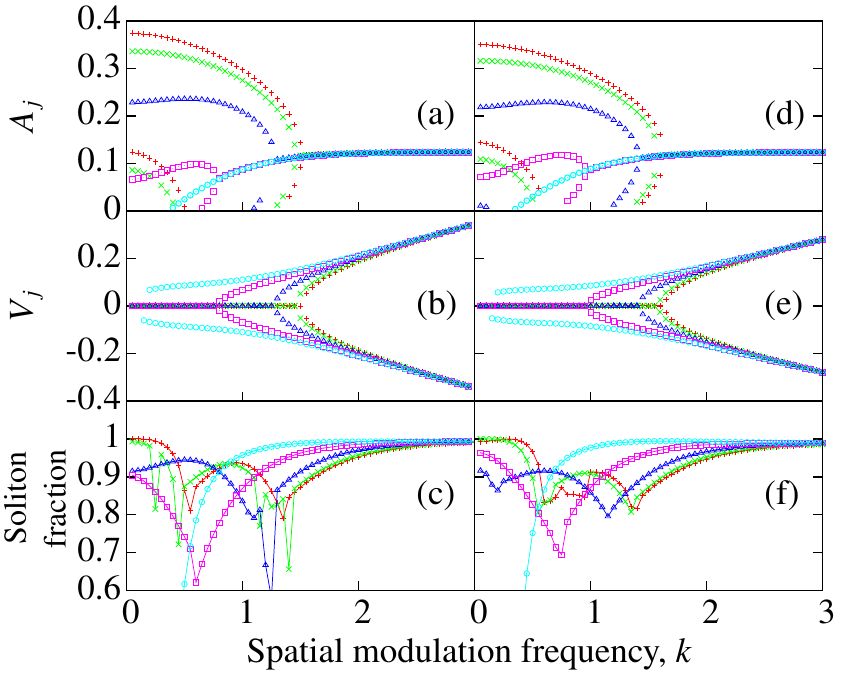}
\caption{(color online). The structure of a multi-soliton pulse. The panels
show the soliton amplitudes $A_j$, velocities $V_j$, and fractions associated
with the initial condition $\psi_0(x)$ \eqnrefp{multi_soliton_solutions} in the
NLSE \eqnrefp{nlse}, computed using a numerical scattering transform
\cite{boffetta_osborne_jcp_1992}, as a function of spatial modulation frequency
$k$. Panels (a--c) correspond to $\alpha =2$ and (d--f) to $\alpha=2.2$.
Relative phases are $\Phi =0$ (\red{$+$}), $\pi/4$ (\green{$\times$}), $\pi/2$
(\blue{$\bigtriangleup$}), $3\pi/4$ (\magnta{$\square$}), $\pi$
(\cyan{$\circ$}). Soliton fraction is the ratio of the combined norm of the
constituent solitons, $\sum_j 2A_j$, to the total norm $\int_{-\infty}^{\infty}
\vert \psi_0(x) \vert^2 dx$ \cite{NoteAmp}. In the limit $k \rightarrow
\infty$, when $\alpha = 2$, $A_j \rightarrow 1/8$ [$\sum_j 2A_j \rightarrow
\int_{-\infty}^{\infty} | \psi_0(x) |^2 dx \rightarrow 1/2$], and $V_j
\rightarrow \pm k/8$ \cite{kodama_hasegawa_1991,
afanasjev_vysloukh_j_opt_soc_am_b_1994}.}
\label{dst_figure}
\end{figure}

Neglecting the axial trapping (setting $\omega = 0$) the 1D GPE
\eqnrefp{1dgpe_soliton_units} reduces to the NLSE \eqnrefp{nlse}, and the
ground state of the BEC before the change in scattering length,
$\psi_\alpha(x)$, is a single, stationary bright soliton \cite{NoteAmp}. After
density modulation
\begin{equation}
\psi_0(x) = \frac{1}{2\alpha} \sech{\frac{x}{2\alpha^2}} \cos
\left(\frac{kx}{2\alpha^2} + \frac{\Phi}{2} \right).
\label{multi_soliton_solutions}
\end{equation}
 
Solutions of the NLSE for this initial condition
\eqnrefp{multi_soliton_solutions} are well-known in the context of nonlinear
optics \cite{satsuma_yajima_1974, kodama_hasegawa_1991,
afanasjev_vysloukh_j_opt_soc_am_b_1994} . The case $k=0$ has been studied
analytically by Satsuma and Yajima \cite{satsuma_yajima_1974} using the inverse
scattering transform (IST) \cite{zakharov_shabat_1972_russian}: for integer
$\alpha=J$, \eqnreft{multi_soliton_solutions} consists of a bound state, or
multi-soliton pulse, of $J$ solitons with unequal amplitudes $A_j$ and zero
velocity ($V_j = 0$). For non-integer $\alpha =J+\beta$,
\eqnreft{multi_soliton_solutions} consists of $J$ solitons plus radiation, with
the norm of the soliton component given by $\sum_j 2A_j$
\cite{satsuma_yajima_1974,NoteAmp}. The modulated case (general $k$) has been
considered both analytically and numerically by Kodama and Hasegawa
\cite{kodama_hasegawa_1991} and Afanasjev and Vysloukh
\cite{afanasjev_vysloukh_j_opt_soc_am_b_1994}.  \figreftcap{dst_figure}{} shows
how the modulation alters the character of a two soliton pulse
($\alpha\gtrsim2$): beyond a certain threshold value of $k$ the pulse
``splits'' into two solitons with equal amplitudes, opposite velocities, and
relative phase $\Phi$, plus a negligible radiation component. Crucially,
control of the modulation corresponds to control over the relative velocity and
phase of a pair of generated bright solitons.

\begin{figure}[t]
\begin{center}
\includegraphics{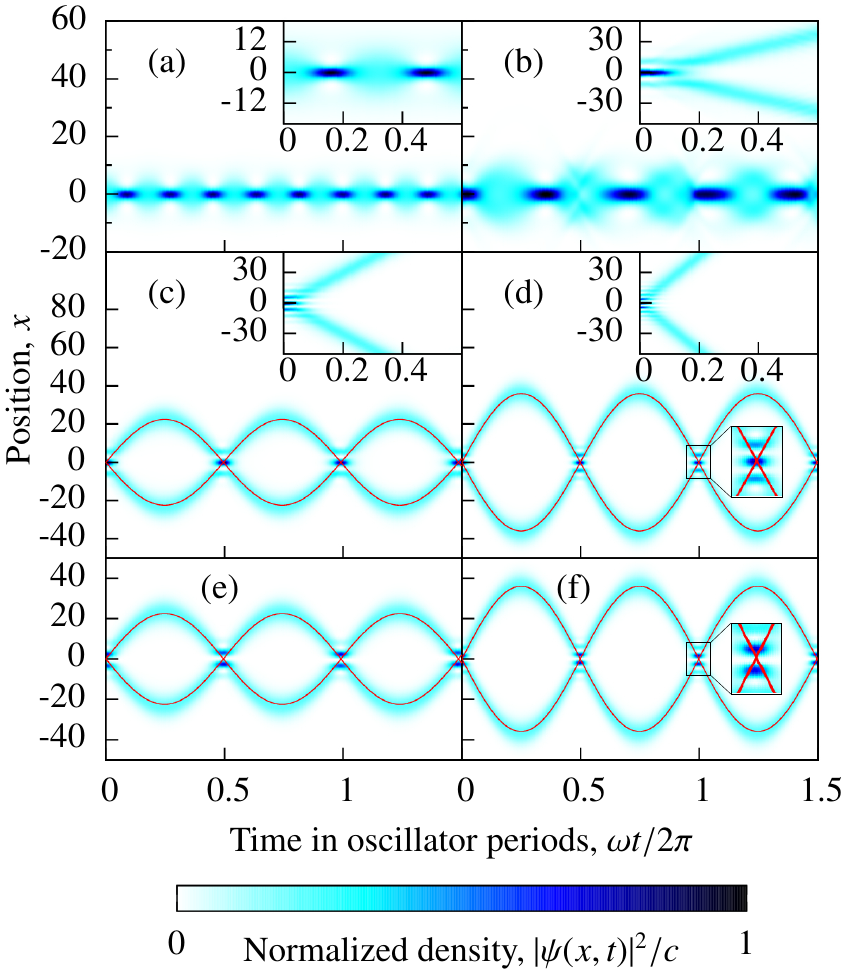}
\end{center}
\caption{(color online). Generation of \BSWs with controlled relative phase via
the interference protocol, in the quasi-1D limit. Panels (a--f) show the
evolution of 1D GPE with trap frequency $\omega=0.02$ [$\omega=0$ inset in
(a--d)] and initial condition $\psi_0(x)$ for $\alpha = 2$, $\Phi=0$ and $k=0$
(a), $k=2$ (b), $k=4$ (c), $k=6$ (d), and $\Phi=\pi$ and $k=4$ (e), $k=6$ (f),
computed using a pseudospectral split-step method. Particle model
\cite{martin_etal_prl_2007, martin_etal_PRA_2008} \BSW trajectories, for
effective masses and velocities obtained from the numerical scattering
transform of $\psi_0(x)$, are overlaid as lines in (c--f). Panels (e) and (f)
reproduce (c) and (d) for the case $\Phi=\pi$ to show the difference in
collision profile. The density (color) axes are normalized by $c=0.35$ (inset
$c=0.25$) in (a) and $c=0.12$ (inset $c= 0.07$) in (b--f).}
\label{1d_figure}
\end{figure}

A similar correspondence exists in the presence of axial trapping ($\omega>0$).
In this regime the 1D GPE \eqnrefp{1dgpe_soliton_units} has no soliton
solutions so we study the dynamics of initial condition $\psi_0(x)$
numerically, concentrating, for simplicity, on the case $\alpha=2$ (other cases
$\alpha \gtrsim 2$ are similar except for a slightly altered relationship
between $k$ and the resulting soliton speed). A pair of equal amplitude \BSWs
are generated with relative phase $\Phi$ and velocities controlled by $k$
\figrefp{1d_figure}{}. The axial trap confines the outgoing \BSWs and causes
subsequent re-collisions at the trap center, and the relative phase upon
re-collision is always identical to the original imposed relative phase
\cite{martin_etal_PRA_2008}. The \BSWs remain highly soliton-like: the density
profile during \BSW collisions is similar to that for bright solitons
\cite{gordon_opt_lett_1983} \figrefp{1d_figure}{(d,f)}, the \BSW trajectories
are well described by a particle model \cite{martin_etal_prl_2007,
martin_etal_PRA_2008, poletti_etal_prl_2008} \figrefp{1d_figure}{(c--f)}, and
the \BSWs are stable against their mutual collisions, insofar as they retain
their form for a sufficiently large number of collisions that atom losses,
unaccounted for in the GPE, would be the lifetime-limiting factor in an
experiment.

\begin{figure}[t]
\begin{center}
\includegraphics{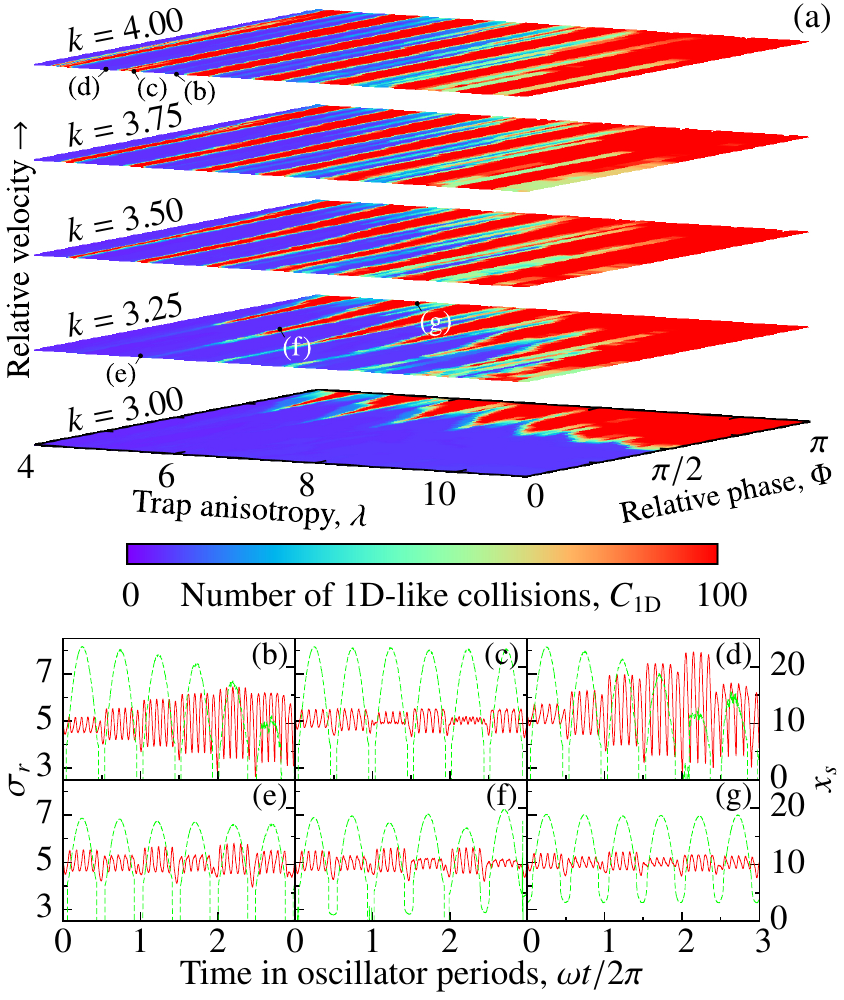}
\end{center}
\caption{(color online). Stability of \BSW collisions in 3D. Panel (a) shows
the number of 1D-like \BSW collisions, $C_{\rm 1D}$, as a function of $k$,
$\Phi$, and $\lambda$ \cite{Note3}.  Effective trap frequency $\omega=0.02$ and  $\alpha
=2$. Also shown is the evolution of the positively displaced \BSW position,
$x_s$ (dashed green line, right vertical axis), and the full width at half
maximum of the integrated radial density distribution, $\sigma_r$ (solid red
line, left vertical axis), at the indicated points on the $k=4$ (b--d) and
$k=3.25$ planes in (a). The computation leverages the radial symmetry of the
problem, using a pseudospectral split-step method in 2D cylindrical
coordinates.}
\label{3d_figure}
\end{figure}

Moving beyond the quasi-1D regime, we can generate pairs of 3D \BSWs with
controlled velocity and relative phase using the same method.  Dynamics in the
radial directions can affect the stability of the \BSWs, however; in certain
cases this drastically reduces the number of collisions for which they retain their
form. Parameterizing the quasi-1D-to-3D transition by the trap anisotropy
$\lambda = \omega_r / \omega_x$, we write the 3D GPE as
\begin{equation}
i\frac{\partial \psi(\mathbf{r},t)}{\partial t} = \left[-\frac{\nabla^2}{2} +
V (\mathbf{r}) - \frac{2\pi}{\lambda \omega} \vert \psi(\mathbf{r},t)
\vert^2 \right] \psi(\mathbf{r},t), \label{3dgpe_soliton_units} 
\end{equation} 
where we use soliton variables \cite{Note2}, and $V (\mathbf{r}) = \omega^2 [ x^2 +
\lambda^2 (y^2 + z^2)]/2$. We again study the dynamics of the \BSWs
numerically, quantifying their stability against collisions in terms of their
positions and maximum integrated axial densities at the point of maximum
separation --- this being much easier to measure, on typical experimental
scales, than the exact density profile during the collision.
\figreftcap{3d_figure}{(a)} shows how the number of 1D-like collisions $C_{1D}$
(taken to be those where the positions and maximum integrated densities of the
BSWs subsequently return to within 75\% of their original values) depends on
velocity, relative phase, and trap anisotropy. We term these collisions 1D-like
because all collisions of quasi-1D \BSWs satisfy these criteria ($C_{\rm 1D}
\rightarrow \infty$).

As expected, \figreft{3d_figure}{(a)} shows that $C_{\rm 1D}$ is strongly
dependent on the relative phase at low velocity, with the \BSWs being most
stable around $\Phi=\pi$ outside the quasi-1D regime \cite{Note3}. At higher
velocity this phase-dependence weakens and the quasi-1D regime is
reached at lower anisotropy. \figreftcap{3d_figure}{(a)} also reveals a
previously unobserved feature: $C_{\rm 1D}$ shows a strong, oscillatory
dependence on the anisotropy at all velocities. This dependence arises from the
\BSWs being broken up by the transfer of energy to radial oscillations
\figrefp{3d_figure}{(b--d)}. These oscillations are started by the abrupt
change in scattering length, and subsequently amplified by collisions if the
\BSWs collide when their radial width is close to its oscillatory maximum. The
frequency of the radial oscillations is primarily determined by $\omega_r$;
this leads to the observed oscillations of $C_{\rm 1D}$ as a function of
$\lambda$. The amplifying effect of collisions decreases with the \BSW
velocity, and at low velocity a phase-dependent amplification of the radial
oscillations emerges \figrefp{3d_figure}{(e--g)}, which we attribute to the
higher densities at the point of collision when $\Phi=0$ delivering a larger
``kick'' than when $\Phi=\pi$.  However, for intermediate phases
symmetry-breaking population transfer \cite{parker_etal_JPB_2008,
khaykovich_malomed_pra_2006} during collisions also contributes to the
reduction in $C_{\rm 1D}$ \figrefp{3d_figure}{(f)}. Within the GPE description,
\figreft{3d_figure}{} represents a comprehensive prediction of the \BSW dynamics
resulting from our splitting protocol. Experimental observation of the dynamics
we predict would support the validity of the GPE description of \BSWs and, in
the case of the oscillatory dependence of $C_{1\rm D}$ on $\lambda$, open the
possibility of controlling the \BSW lifetime directly. 

To conclude, we have proposed an experiment that produces a pair of \BSWs with
controlled relative phase and velocity in a harmonically trapped atomic BEC,
and we have analyzed the subsequent collisions of these \BSWs using the GPE. In
the quasi-1D regime the \BSWs are highly soliton-like and stable against their
re-collisions. In the fully 3D regime, we confirm that the collisional
stability of the \BSWs depends on their relative phase and velocity, and
demonstrate for the first time a strong oscillatory dependence on the trap
anisotropy.  The presence, or absence, of these effects in experiments provides
a direct test of whether experimentally observed atomic BSWs can be described
in terms of a coherent effective single-particle wavefunction, propagated by
the GPE.

We thank N. G. Parker, S. A. Wrathmall, and P. M. Sutcliffe for many
stimulating discussions, and UK EPSRC (Grant No. EP/G056781/1), the Royal
Society (SLC), and Durham University (TPB) for support.


\begin{thebibliography}{10}
\bibitem{NoteAmp}%
  \BibitemOpen
  \bibinfo {note} {Note, for any amplitude $A$, that $A\,\sech{Ax}$ is a
  stationary soliton solution to \eqnreft{nlse}, with norm $2A$.}%
  \bibAnnoteFile{Stop}{NoteAmp}%
\end{thebibliography}
%

\end{document}